\documentclass[a4paper,11pt]{article}
\usepackage{jinstpub} 
\usepackage{lineno}
\usepackage{siunitx}
\usepackage{subcaption}

\title{\boldmath CosmoLink: Portable Coincidence Detector for On-Site Muon Flux Measurement}

\author[1]{Yuvaraj Elangovan,\note{Corresponding author.}}
\affiliation{University of Pittsburgh}

\emailAdd{yue8@pitt.edu}

\abstract{
The CosmoLink project aims to develop low cost, low power, long range and portable form factor particle detectors for experimental and educational research on Cosmic Rays. Use of these detectors distributed across locations will be used for setup small scale telescope array experiments. Each detector unit in CosmoLink project consists of two plastic scintillators kept in stacked geometry to look for coincidence induced by cosmic ray particles. The scintillators are directly coupled with Silicon photomultipliers (SiPMs) for efficient light guiding. Each readout channel is equipped with an amplifier, discriminator and a peak-detect-and-hold circuit. Upon successful coincidence, the peak amplitude is digitized using Analog to Digital Converter (ADC). The coincidence count and ADC data are acquired using a low cost microcontroller. Multiple detectors can be grouped together like swarms and arranged in different geometries to collect data jointly. The acquired data is wirelessly transmitted to a central server with GPS Coordinates and other sensor data.}


\begin{document}
\maketitle
\flushbottom

\section{Introduction}
\label{sec:intro}

The objectives of the CosmoLink project are the design of a low-cost small scale telescopic array to acquire real-time cosmic shower data in distributed locations, the creation of a framework for building small-scale particle detectors that can be used in educational and research settings and the creation and maintenance of an online forum named CosmoLink with official web server and data sharing capabilities for academic research. This project draws its inspiration from the Cosmic-Watch program developed by MIT \cite{cosmicwatch} but incorporates significant enhancements, including improved coincidence measurement capabilities, a more adjustable detector size, faster readout electronics and long distance wireless data transfer. Each detector unit in CosmoLink is supported by low noise, low power and compact high speed Data acquisition system. Plastic Scintillators are used as active detectors for counting cosmic ray muon events. SiPM and associated readout electronics composed of amplifier, discriminator and peak hold circuit is used for signal processing from the scintillators. Coincidence triggering, facilitated by the central microcontroller (Raspberry Pi Pico W) ensures that only events detected by both scintillators simultaneously are considered valid. This mechanism reduces false positives and enhances the reliability of muon flux measurements. Once a valid trigger occurs, the acquired data, including coincidence counts, peak amplitudes and timestamps are transmitted wirelessly to a server for storage and analysis. The Raspberry Pi Pico W is a low-cost, high-performance microcontroller board based on RP2040 with flexible digital interfaces.  A dedicated GPS, Inertial Measurement Unit (IMU) and Ambient sensor is interfaced with the Pico to provide real-time telemetry along with particle data. The unit is powered by rechargeable battery as well as from mains power supply. Dedicated RF transceivers are used for long range communication to the server. The detector can be easily transported and deployed in diverse locations from remote field sites to laboratory settings without requiring huge logistics and access to mains power. This setup allows for accurate detection and measurement of muon flux in various environments making it a tool for research, monitoring and experimentation. Currently two detector units and one server unit are under development for prototyping. This paper covers the design of the prototype module of this project.

\section{Detector Geometry}
\label{sec:geo}
\begin{figure}[htbp]
\centering
\includegraphics[width=.7\textwidth]{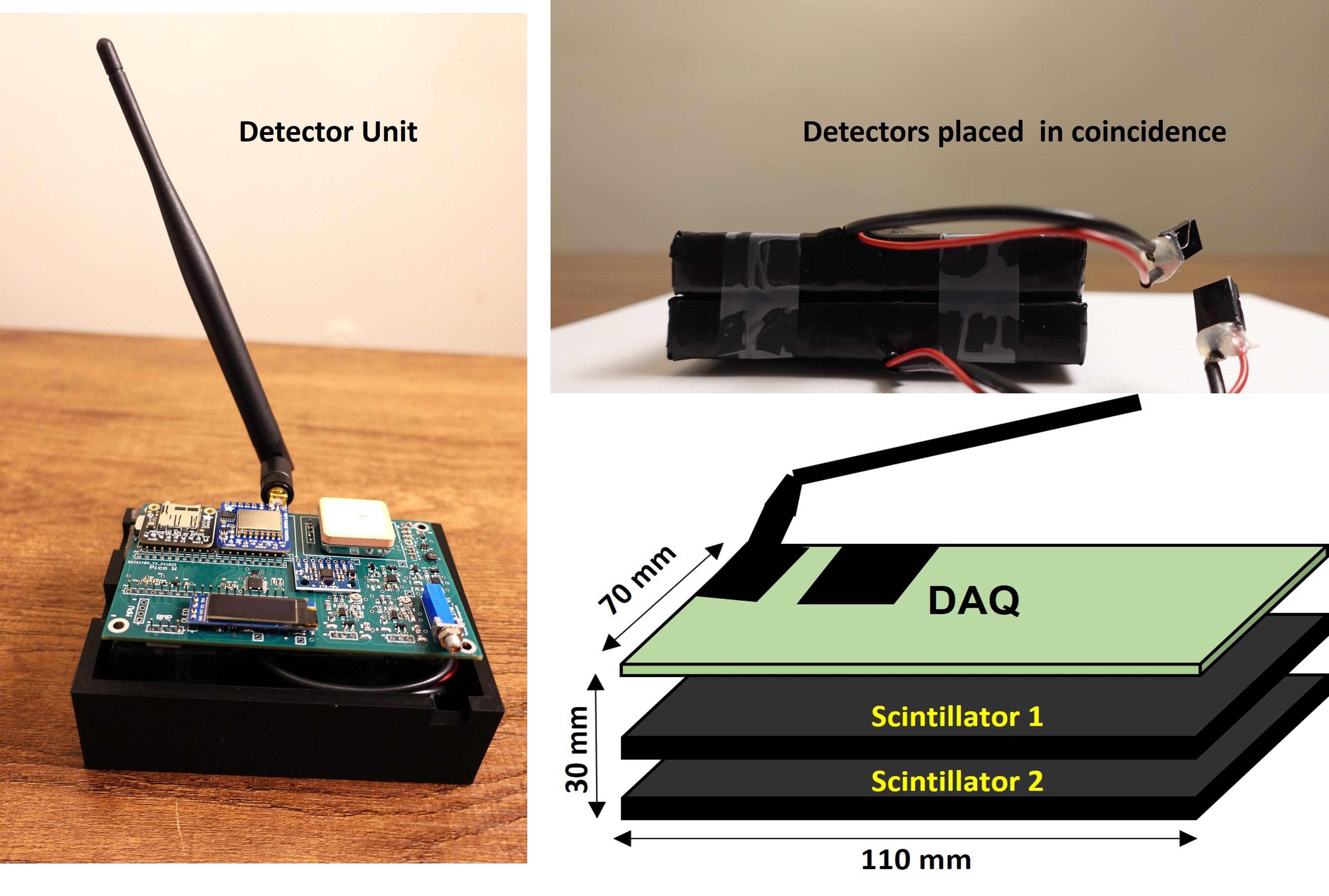}
\caption{Prototype detector unit and proposed design geometry of the coincidence detector unit.\label{fig:1}}
\end{figure}

The CosmoLink Detector Unit is a compact, small-sized particle detector featuring two plastic scintillators. The Data Acquisition (DAQ) printed circuit board measures $110\, \text{mm} \times 70\, \text{mm}$. Two plastic scintillators, each $100\, \text{mm} \times 50\, \text{mm}$ are stacked vertically beneath the DAQ board to enable coincidence measurements as shown in figure~\ref{fig:1}. The detector unit has an overall height of approximately $30\, \text{mm}$. The SiPMs are mounted on a $20\, \text{mm} \times 20\, \text{mm}$ carrier board, which can be coupled with the scintillators. First the scintillators are wrapped with reflective sheets to enhance light collection, followed by a layer of black tape to shield them from external light. Provisions are included for connecting an antenna and a battery, ensuring portability and communication capability. The entire unit is enclosed within a plastic chassis for protection and durability. The enclosure protects the  components and helps in handling. The DAQ is placed on top of the two detectors using metal studs. Slots were made to USB cable connection for powering and readout. 

\section{Trigger and Data Acquisition}
\label{sec:tdaq}
\begin{figure}[htbp]
\centering
\begin{subfigure}[c]{0.45\textwidth}
\includegraphics[width=\linewidth]{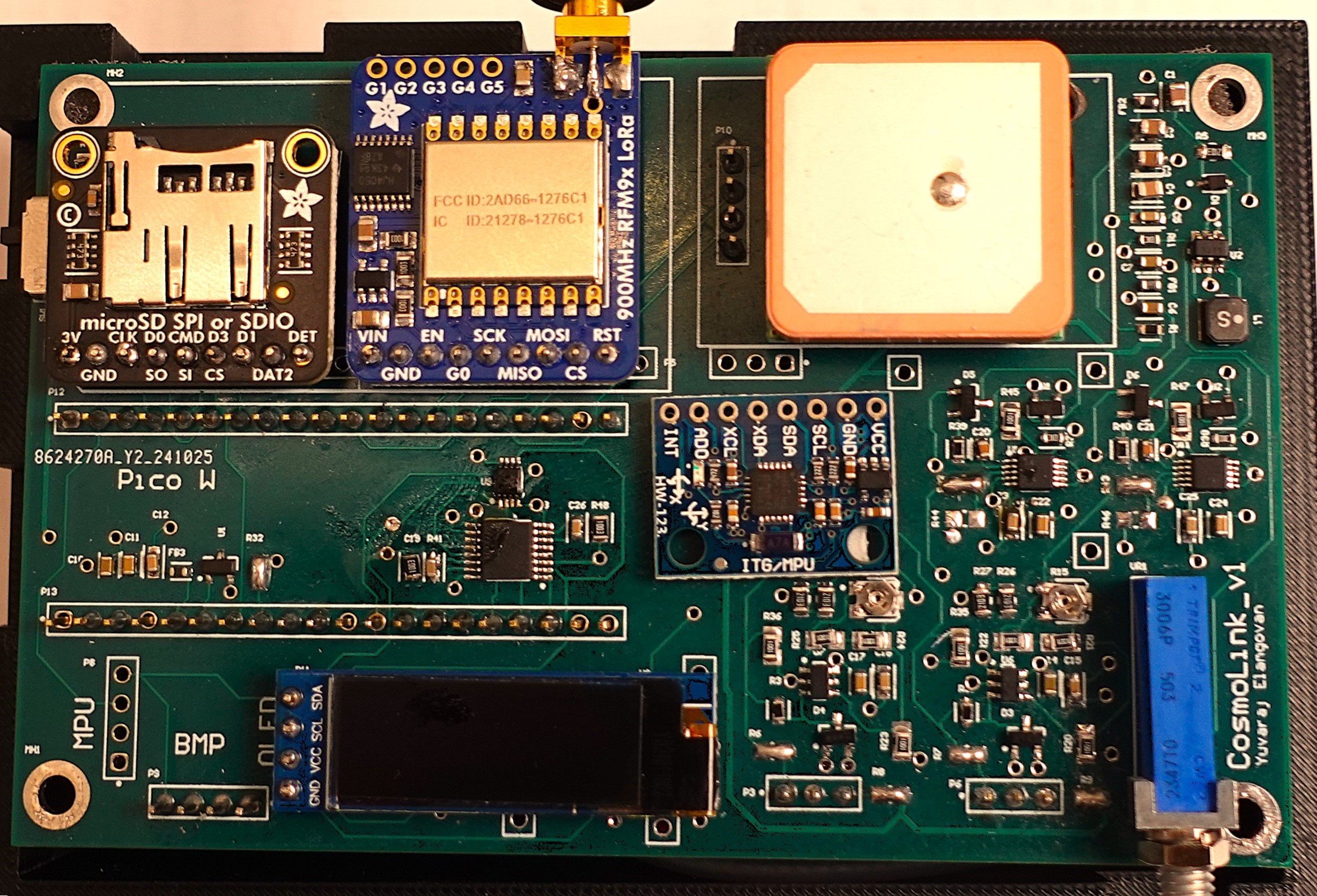} 
\caption{DAQ Module Top.}
\label{fig:3a}
\end{subfigure}\hfill  
\begin{subfigure}[c]{0.48\textwidth}
\includegraphics[width=\linewidth]{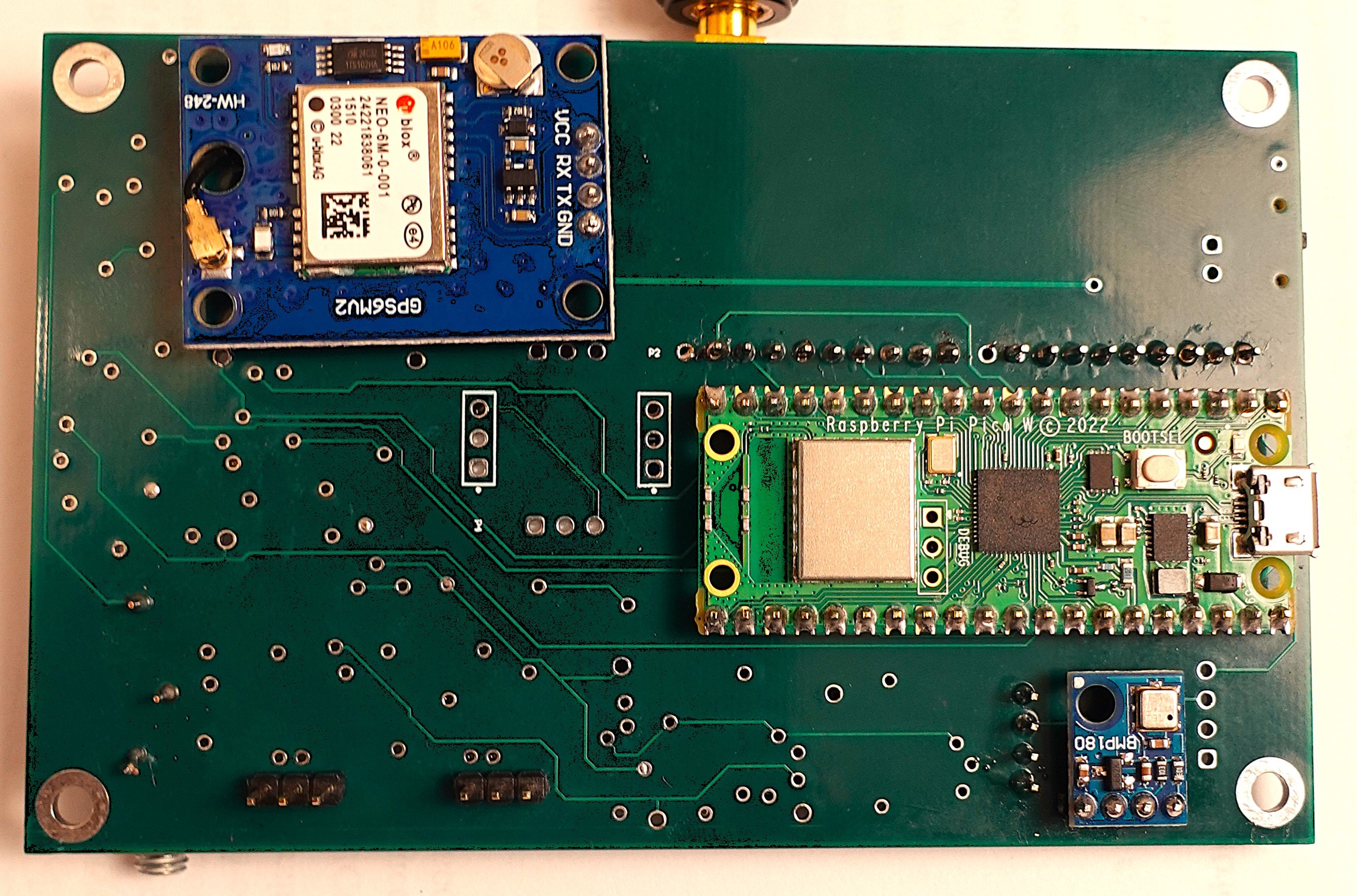}
\caption{DAQ Module Bottom.}
\label{fig:3b}
\end{subfigure}
\caption{DAQ Mother Board.\label{fig:3}}
\end{figure}

The detection process begins when the scintillators capture incident muons, producing scintillation light that is collected by SiPMs (MICROFC-60035-SMT) \cite{sipm}. Each SiPM is biased at 30V using a custom-built adjustable DC-DC converter which can be tuned from 27V to 33V. Additionally, the SiPM bias includes in-situ temperature compensation to maintain stability. Each scintillator has a dedicated readout channel,
\begin{figure}[htbp]
\centering
\includegraphics[width=1\textwidth]{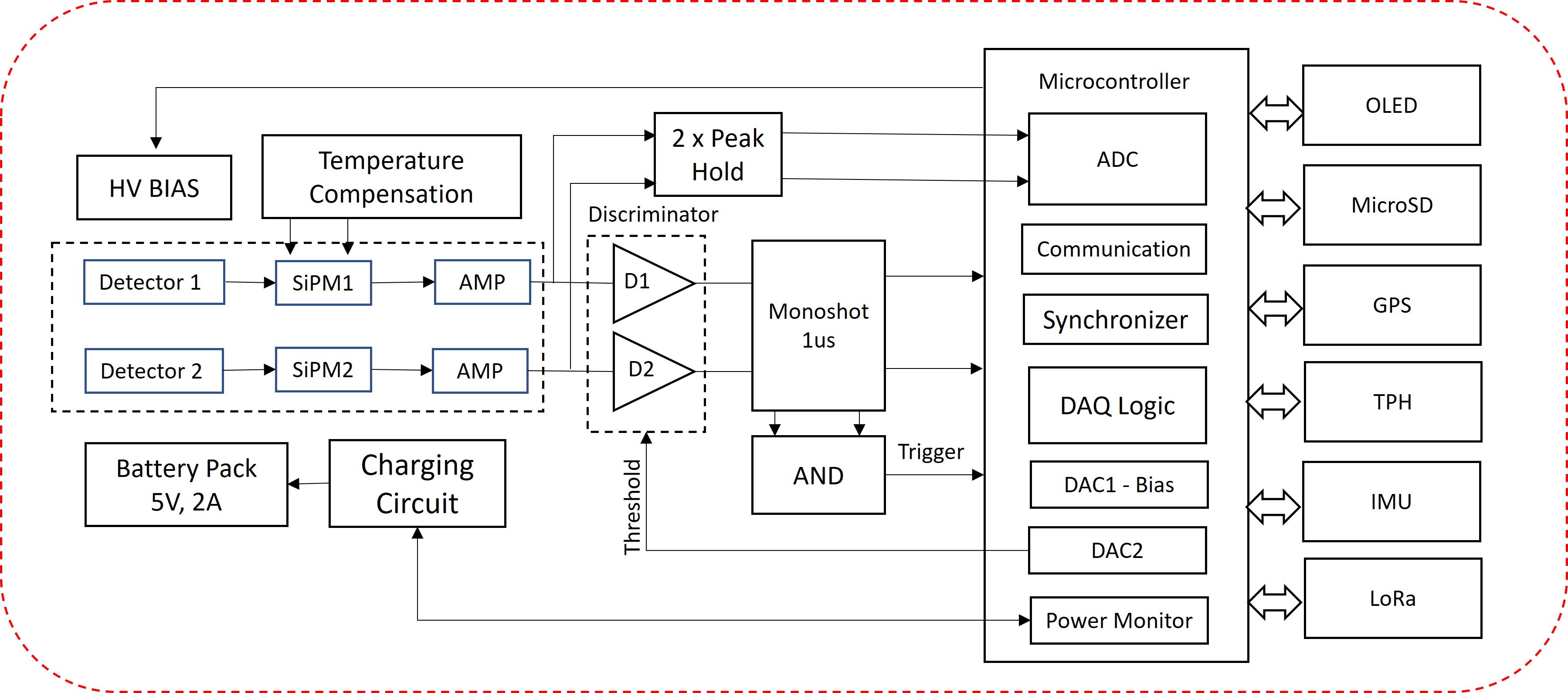}
\caption{DAQ Schematic.\label{fig:2}}
\end{figure}
equipped with an amplifier, discriminator and peak-detect-and-hold circuit as shown in figure~\ref{fig:2}. The preamplifier gain is set to 30 and the discriminator threshold is carefully selected to avoid SiPM dark noise. At the core of the DAQ system is a Raspberry Pi Pico W microcontroller \cite{pico} which handles readout and communication. Two ADC channels on the Pico W are assigned to the two scintillators. The peak amplitude of the signal is held for up to $50\, \text{$\mu$s}$. Upon a muon event detection, the peak value is digitized by the ADC. A fast trigger signal is generated using an AND gate, which increments a counter and initiates ADC conversion of the peak-held value. Once the Pico completes the ADC reading and counting, the peak circuit is reset. The Raspberry Pi Pico W's GPIOs are used to receive the trigger as an interrupt while the peak-held signals are connected to analog input pins. The top and bottom side of DAQ module are shown in figure~\ref{fig:3}. For additional measurements an MPU6050 IMU (Inertial Measurement Unit) sensor is integrated
to provide relative angular data and a BMP180 sensor records real-time temperature and pressure. A NEO 6M GPS module is used for accurate time-stamping of events with geolocation. One of the device’s key features is its portability, powered by a battery or a 5V USB supply, making it easily deployable in various environments, from remote field locations to lab settings without requiring mains power. The battery voltage is monitored through an ADC pin on the Pico, ensuring continuous operation.

\section{Data Framing and Logging}
\label{sec:event}
During an event the coincidence trigger generates an interrupt that initiates an event Interrupt Service Routine (ISR) in the Pico microcontroller. Within the ISR the analog read function retrieves ADC values from both peak-held signals and increments the individual scintillator trigger counters as well as the coincidence trigger counter. The trigger dead-time is around $50\, \text{$\mu$s}$. An event flag is set in the ISR while exiting the routine. In the main loop, this event flag is checked; if it is high, peripheral readout is initiated. At this point, the BMP180 sensor and MPU6050 are read using the I2C protocol, while GPS coordinates are obtained via UART. A data frame shown in table~\ref{tab:1} is constructed to include counter data, peak values, temperature, pressure, GPS coordinates and angular data. This data frame is then saved to the microSD card. Additionally, a periodic timer triggers the transfer of accumulated data to the server.

\begin{table}[htbp]
\centering
\caption{Event Data Format.\label{tab:1}}
\smallskip
\begin{tabular}{|l|l|}
\hline
Datatype &Description \\
\hline
Peak1  & Peak value of Scintillator1\\
Peak2  & Peak value of Scintillator2\\
Count1 & Scintillator1 counter value\\
Count2 & Scintillator2 counter value\\
Coincidence & Coincidence trigger counts\\
Temperature & Temperature on-site\\
Pressure & Pressure in mbar\\
Altitude & Altitude in meters\\
GPSdata0  & latitude string\\
GPSdata1  & longitude string\\
Angle  & angle in deg\\
\hline
\end{tabular}
\end{table}

\section{Wireless Detector Network}
\label{sec:wdn}
\begin{figure}[htbp]
\centering
\includegraphics[width=.7\textwidth]{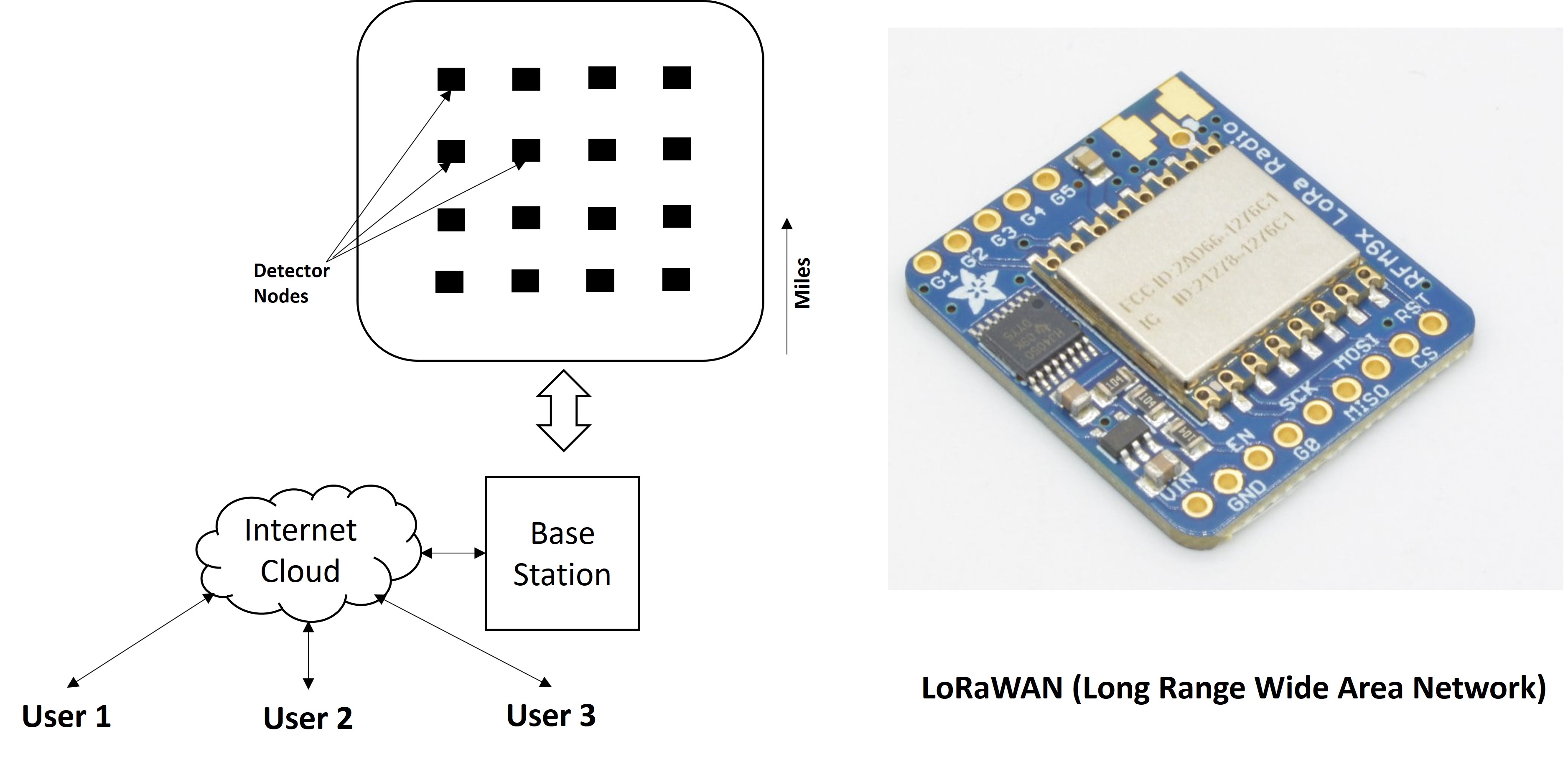}
\caption{Wireless Detector Network (WDN) and Adafruit RFM95W LoRa Radio Transceiver.\label{fig:4}}
\end{figure}

Multiple detectors can be deployed across geographical locations covering large area configurations allowing collective measurement capabilities. This setup transforms the device into a large-area detector providing monitoring and analysis of muon flux patterns across different spatial scales. To support this distributed detector network LoRaWAN \cite{lora} is used for communication between the detectors and the central server as shown in figure~\ref{fig:4}. With a typical range of 5 to 15 kilometers, LoRaWAN is well-suited for wireless detector networks. Additionally, the Pico W’s built-in WiFi and Bluetooth can be utilized for service and maintenance purposes. A web-server will be used for real-time monitoring of detector network.

\section{Results}
\label{sec:result}
\begin{figure}[htbp]
\centering
\includegraphics[width=.9\textwidth]{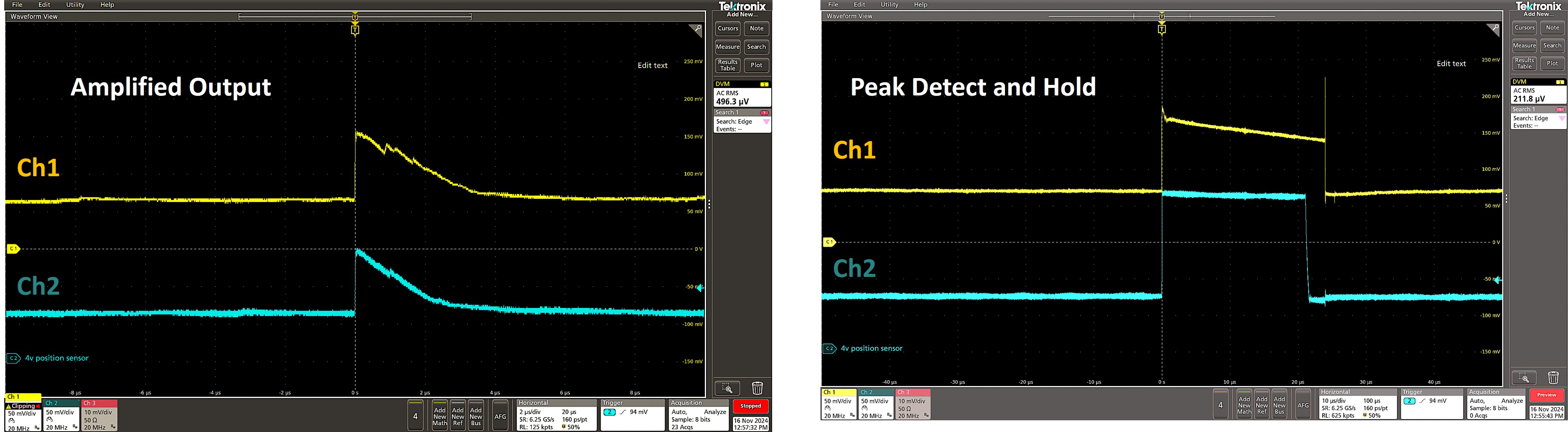}
\caption{DAQ signals observed during an coincident event.\label{fig:5}}
\end{figure}

\begin{figure}[htbp]
\centering
\includegraphics[width=.7\textwidth]{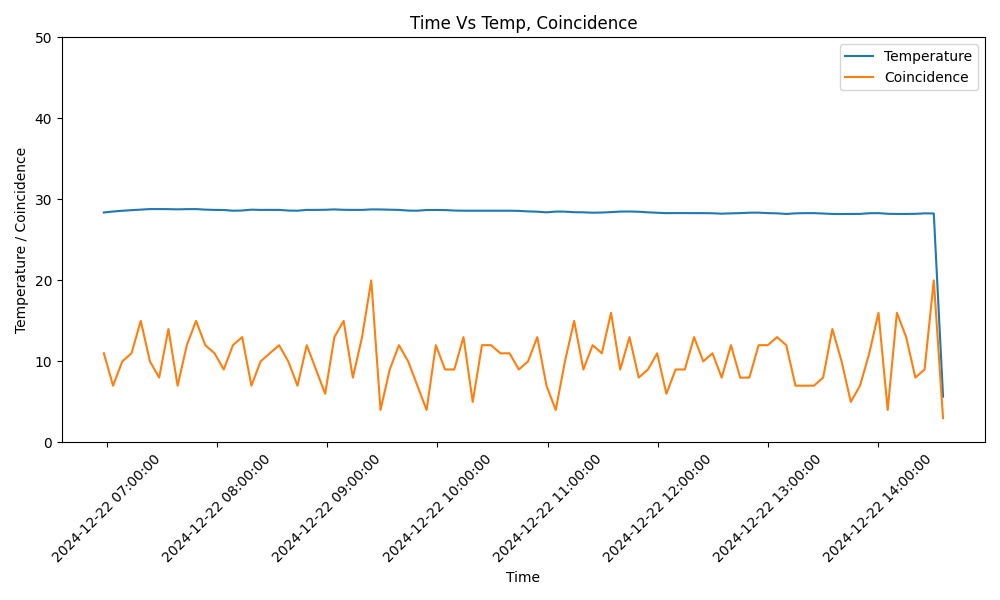}
\caption{Coincidence Vs Temperature over time.\label{fig:6}}
\end{figure}

The prototype detector unit was initially bench-tested in the laboratory, with signal probing conducted at various stages. The coincidence between the amplified pulses and peak-held signals was analyzed using an oscilloscope, as shown in figure~\ref{fig:5}. To evaluate the detector module's performance and the readout electronics under continuous operation, data from the detector unit was logged over an extended period. The coincidence counts and temperature readings in degree celsius were recorded over time, with data logged every 10 seconds and averaged over 15 minutes for plotting, as shown in figure~\ref{fig:6}.

\section{Conclusion}
Currently two units of the CosmoLink detector and a server unit have been developed and are undergoing testing. Initial measurements indicate relatively low count rates, which can potentially be improved by adjusting the amplifier gain and discriminator threshold. Additionally, a Geant4-based simulation is planned to analyze the detector material and size requirements to enhance performance. The existing DAQ design supports larger scintillators, provided the trigger rate stays below 20 kHz and SiPM-to-scintillator coupling is optimized. We are currently testing the proof of concept using these prototype units to demonstrate their potential and secure funding. The next phase of the project aims to develop a multi-node detector network involving collaboration with schools and universities.


\end{document}